\documentclass[aps, prb, reprint, superscriptaddress]{revtex4-2}

\usepackage{graphicx} 
\usepackage{braket}
\usepackage{physics}
\usepackage{amsmath}
\usepackage{amssymb}
\usepackage{mathtools}
\usepackage{bbold}
\usepackage{hyperref}
\usepackage[capitalize]{cleveref}
\usepackage{xcolor}
\usepackage{comment}

\renewcommand{\Im}{\mathrm{Im}}

\begin{document}
\author{Ammar Kirmani}
\affiliation{Theoretical Division, Los Alamos National Laboratory, Los Alamos, New Mexico 87545, USA}

\author{Andrew A. Allocca}
\affiliation{Physics Department, City College of the City University of New York, New York 10031, USA}

\author{Jian-Xin Zhu}
\affiliation{Theoretical Division, Los Alamos National Laboratory, Los Alamos, New Mexico 87545, USA}
\affiliation{Center for Integrated Technologies, Los Alamos National Laboratory, Los Alamos, New Mexico 87545, USA}

\author{Armin Rahmani}
\affiliation{Department of Physics and Astronomy and Advanced Materials Science and Engineering Center, Western Washington University, Bellingham, Washington 98225, USA}

\author{Sriram Ganeshan}
\affiliation{Physics Department, City College of the City University of New York, New York 10031, USA}
\affiliation{Physics Program, Graduate Center of City University of New York, New York 10031, USA}

\author{Pouyan Ghaemi}
\affiliation{Physics Department, City College of the City University of New York, New York 10031, USA}
\affiliation{Physics Program, Graduate Center of City University of New York, New York 10031, USA}

\title{Hall viscosity of the Laughlin state on noisy quantum computers}
\begin{abstract}
Hall viscosity is a quantized nondissipative stress response of a fractional quantum Hall (FQH) fluid to adiabatic geometric deformations. Despite strong theoretical interest, its experimental observation in the FQH state has remained elusive, making it a demanding target for realization on current {IBM quantum hardware}. In this letter, we employ a quasi-one-dimensional model of an FQH state coupled to a background metric to probe the geometric response under a metric quench. We design and implement a quantum-circuit protocol that realizes a Hilbert-space-truncated version of the model and extracts the Hall viscosity from the geometric response encoded in the wave-function dynamics on the {IBM quantum hardware}. While the truncation prevents us from accessing the fully quantized value of Hall viscosity, the hardware data  {closely track} analytical and numerical predictions within this  regime.
\end{abstract}

\maketitle

\section{Introduction}
Quantum simulation is among the most promising applications of quantum computers, enabling controlled emulation of complex quantum systems that are intractable for classical computation and even difficult to study in actual experiments~\cite{feynman1982simulating,lloyd1996universal}. 
The advent of programmable quantum computing platforms, including superconducting qubits~\cite{Kandala2017}, trapped ions~\cite{Monroe2021}, and neutral atoms~\cite{Browaeys2020}, has enabled remarkable experimental progress. 
These platforms are beginning to demonstrate key milestones, including the realization of spin models, lattice gauge theories, and strongly correlated quantum phases on current noisy intermediate-scale quantum (NISQ) devices~\cite{georgescu2014quantum,bernien2017probing,klco2018quantum,ebadi2021quantum}. 
Together, these advances benchmark existing hardware and open new avenues for exploring strongly interacting quantum matter, dynamical phenomena, and emergent many-body behavior beyond readily established frameworks ~\cite{altman2021quantum,daley2022practical,Koh2024,10.21468/SciPostPhys.15.4.170,Shen2025}.

Among strongly correlated systems, fractional quantum Hall (FQH) states stand out as an ideal frontier for quantum simulation. FQH states emerge in a two-dimensional electron gas under strong magnetic fields and low temperatures, where electron-electron interactions within nearly dispersionless Landau levels give rise to strongly correlated, topologically ordered phases~\cite{klitzing1980new, laughlin1981quantized, tsui1982two, laughlin1983anomalous}. 
The topological nature of FQH states manifests not only in their quantized electromagnetic response but also in their geometric structure, making them a paradigmatic platform for exploring topological order in quantum simulators.

Recent works, including by three of the present authors~\cite{Rahmani2020, Kirmani2022,kirmani2023}, have demonstrated that NISQ devices can successfully realize and probe FQH-like states.  
In particular, FQH states {in thin torus geometry}~\cite{Nakamura2012,PRXQuantum.6.020321,PhysRevX.5.041003} have proven highly amenable to hardware implementation, enabling the observation of geometric ``graviton'' excitations following a controlled metric quench \cite{Kirmani2022}. 
Further studies have established the fractional statistics of nonlocal anyonic excitations through adiabatic braiding of quasiholes on NISQ devices~\cite{kirmani2023}.

Another property of the FQH state, which has been formidable to realize in conventional experimental measurements, is the Hall viscosity. 
The Hall viscosity is a geometric, nondissipative stress response that directly probes the underlying quantum geometry of FQH states~\cite{avron1995viscosity,avron1998odd, tokatly2007new, tokatly2009erratum, read2009non, haldane2011geometrical, haldane2011self, read2011hall, hoyos2012hall, bradlyn2012kubo, yang2012band, abanov2013effective, hughes2013torsional, hoyos2014hall, laskin2015collective, can2014fractional, can2015geometry, klevtsov2015geometric, klevtsov2017quantum, gromov2014density, gromov2015framing, gromov2016boundary, bradlyn2015low, scaffidi2017hydrodynamic, andrey2017transport, alekseev2016negative, pellegrino2017nonlocal, Pu2020,PRXQuantum.6.020321}. 
While Hall viscosity has proven extremely difficult to observe in conventional condensed-matter experiments (particularly in the quantum fractional Hall regime), its definition in terms of adiabatic deformations of the system’s metric \cite{PhysRevLett.118.146403} makes it promising for investigation on programmable quantum hardware \cite{IBM}. 
In this Letter, we develop a {quantum-hardware} protocol that enables direct {observation} of the Hall viscosity. 
Although our quantum algorithm requires a Hilbert-space truncation imposed by current {quantum hardware} limitations, the results {closely track} analytical and numerical calculations performed in the same limit. 
Our work thus establishes existing NISQ-era quantum hardware as efficient experimental platforms for probing the difficult-to-measure Hall viscosity of the FQH state.

\begin{figure}[!ht]
    \centering
    \includegraphics[width=0.9\columnwidth]{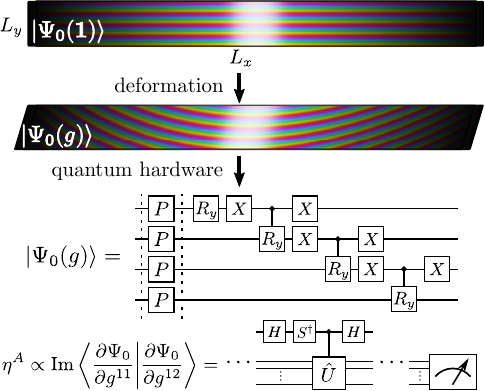}    
    \caption{{Deformation of a thin-torus/cylinder by metric $g$ changes single-particle orbitals and Laughlin state $\ket{\Psi_0(g)}$, which can be prepared on quantum hardware. 
    Circuit operations transpiled to quantum hardware compute the Berry curvature of $\ket{\Psi_0(g)}$ with respect to $g^{11}, g^{12}$ allowing observation of the Laughlin state's Hall viscosity $\eta^A$ via \cref{eq:viscosity}.}}  
    \label{fig:sketch}
\end{figure}

\section{Hall viscosity in FQH state in thin torus limit}
In the absence of kinetic energy within a Landau level, FQH states arise solely from electron-electron interactions. 
When interactions are weak compared to the Landau-level energy gap, the effective Hamiltonian of the FQH state is obtained by projecting the interaction Hamiltonian onto the degenerate subspace of a single Landau level. 

Although Coulomb interactions are inherently long-ranged, the ground states associated with certain short-range interactions encode essential features of FQH states. 
The Trugman-Kivelson pseudopotential $V(\mathbf{r})\propto \nabla^2\delta(\mathbf{r})$ \cite{Trugman1985} is one such example where the exact ground state is the Laughlin wave function\cite{laughlin1983anomalous}. 
By projecting the pseudopotential into the orbitals spanning the lowest Landau level of a toroidal system with dimensions $L_x$ and $L_y$, the interaction Hamiltonian yields an effective one-dimensional model in the later orbital space with long-range interactions of the form~\cite{PhysRevLett.51.605,Yoshioka1983}:
\begin{equation}\label{eq:HYoshioka}
    H=\sum_{\{j_i\}=0}^{N_{\phi}-1}A_{j_1,j_2,j_3,j_4}(g)\,\hat{c}^\dagger_{j_1} \hat{c}^\dagger_{j_2} \hat{c}_{j_3} \hat{c}_{j_4}.
\end{equation} 

The effective interaction potentials are
\begin{multline} \label{eq:matrixA}
    A_{j_1j_2j_3j_4}(g)=\frac{\delta'_{j_1+j_2,j_3+j_4}}{2\mathcal{A}} \\
    \times \sum_{s,t\in \mathbb{Z}}\delta'_{j_1-j_4,t}\left(g^{ij}q_i q_j \right)e^{-\frac{1}{2} \left(g^{ij}q_i q_j\right)}e^{2\pi is\frac{j_1-j_3}{N_\phi}}.
\end{multline}
with $q_x = 2\pi s/L_x$ and $q_y=2\pi t/L_y$, and $i,j$ are $x,y$ (i.e. the two axes of the torus) \cite{yang2012band}.
Here, $N_\phi = \mathcal{A}/(2\pi\ell_B^2)$ is the Landau level degeneracy, with magnetic length $\ell_B$ and system area $\mathcal{A}=L_xL_y$. We consider filling fraction $\nu=1/3$.
The operators $\hat{c}_j$/$\hat{c}_j^\dagger$ destroy/create electrons in the $j$th degenerate state of the lowest Landau level, with indices defined mod $N_\phi$. 
Primed $\delta$ functions similarly indicate this periodicity.
The metric $g$ parameterizes the two-dimensional space of the torus and has two independent components since it is unimodular ($\det g =1$).

\begin{figure}[!t]
    \centering
    \includegraphics[width=8cm]{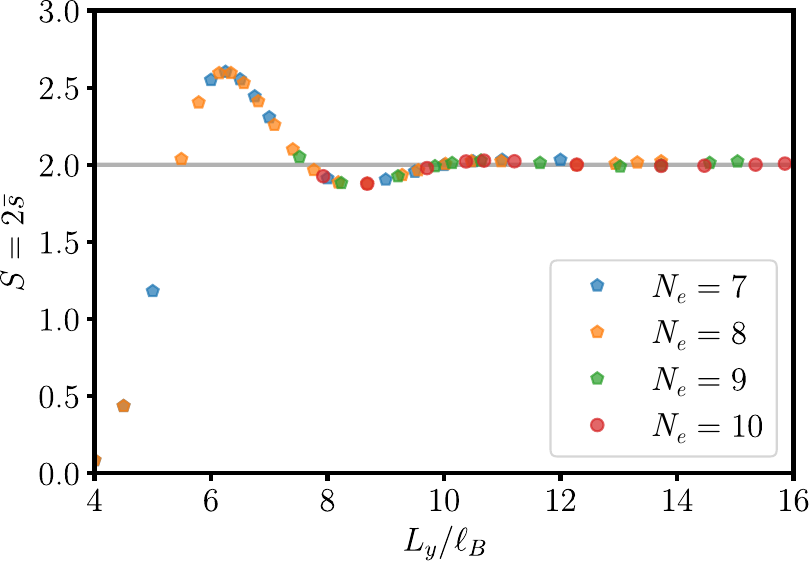}\\
    \includegraphics[width=8cm]{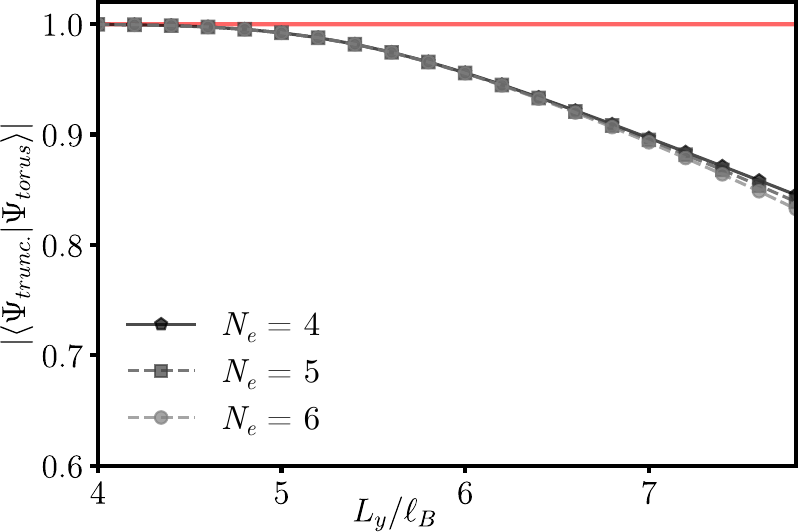}
    \caption{(Top) Spin-shift $S$ obtained from exact diagonalization of \cref{eq:HYoshioka}. 
    The quantization of this quantity verifies that the 1d Hamiltonian and the metric parametrization of the Hall viscosity reproduce known results.
    (Bottom) The overlap of the torus and truncated finite cylinder wave functions.
    {$N_e$ is the number of electrons in the system.} 
    }  \label{fig:1dmodels}
\end{figure}

The Hall viscosity of a FQH fluid is directly related to the Berry curvature associated with adiabatic deformations of the system’s flat geometry~\cite{avron1995viscosity}. 
Such deformations are commonly written with the complex modular parameter $\tau=\tau_1+i\tau_2$, where $\tau_1$ controls the skewness of the torus and $\tau_2$ the squeezing of the plane in $x$ and $y$ dimensions.
In \cref{eq:HYoshioka,eq:matrixA}, we have instead used the metric $g$ with two independent elements, $g^{11}=\abs{\tau}^2/\tau_2$ and $g^{12}=-\tau_2/\tau_2$, which is convenient to implement on quantum {hardware}. 
In these representations, the Hall viscosity is
\begin{equation}
    \eta^A = -\frac{\tau_2^2}{\mathcal{A}}\mathcal{F}_{\tau_1\tau_2}=-\frac{g^{11}}{\mathcal{A}} \mathcal{F}_{g^{11}g^{12}}, \label{eq:viscosity} 
\end{equation}
with Berry curvature $\mathcal{F}_{ab}=-2\Im\braket{\pdv{\Psi_0}{a}}{\pdv{\Psi_0}{b}}$. $a,b$ are either $\tau_1,\tau_2$ or $g^{11},g^{12}$ (see Appendices \ref{app:A},\ref{app:C}, and \ref{app:D}). Equation (\ref{eq:viscosity}) is valid for both torus and cylinder geometries~\cite{Tokatly2009}.  
The derivatives can be evaluated at $\tau_1=0$, $\tau_2=1$ corresponding to $g^{12}=0$, $g^{11}=1$, which is the limit of no deformation. Related quantities used to represent the Hall viscosity are the mean orbital spin per particle $\bar{s} = 2\mathcal{A}\eta^A/N_e$ and the spin shift $S = 2\bar{s}$~\cite{read2011hall}.

The ground state of \cref{eq:HYoshioka} can be obtained using exact diagonalization~\cite{Yoshioka1983}, and the Hall viscosity computed by inserting the ground-state wave function into \cref{eq:viscosity}. 
The result is shown in \cref{fig:1dmodels}(a), demonstrating that the 1d model and the expression for $\eta^A$ in terms of the metric $g$ successfully reproduce the known quantization of this quantity.

To demonstrate the Hall viscosity on the IBM quantum hardware, we need to generate the wave function and apply the unitary transformation associated with the change in geometry. 
The periodic boundary conditions required for torus geometry hardens the realization of the FQH-type wave function on current NISQ hardware. 
Furthermore, the hardware layout of existing quantum {hardware} makes it difficult to implement quantum gates between distant qubits. 
We, therefore, work with a simplified version of the model—one that still captures the essential features of the FQH state while remaining feasible for hardware implementation. 
To this end, we consider the FQH wave function on a finite cylindrical geometry with a truncated Hilbert space. 
Within these approximations, the quasi 1D Hamiltonian of the $\nu=1/3$ Laughlin state~\cite{Nakamura2012}, {which corresponds to the thin torus geometry} is given by
\begin{gather}
     H = \sum_{j=0}^{N_\phi-1}\left[\sum_{k=1}^3V_{k0}n_{j}n_{j+k} + V_{21}c_{j+1}^\dagger c_{j+2}^\dagger c_{j+3}c_j + \mathrm{h.c}\right] \label{eq:Htrunc}\\
    V_{km} \propto (k^2-m^2) e^{-2\pi^2\ell_B^2(k^2+m^2 - 2i g^{12}mk)/(g^{11}L_y^2)} \label{eq:Vkm},
\end{gather}
where the electron operator indices are \emph{not} taken mod $N_\phi$.
The relationship between the torus Hamiltonian [\cref{eq:HYoshioka}] and the truncated Hamiltonian on a finite cylinder [\cref{eq:Htrunc}] is demonstrated in Appendix \ref{app:B}.
Figure \ref{fig:1dmodels}(b) shows that the ground states for these effective Hamiltonians have overlap $\gtrsim0.95$ for $L_y/\ell_B \lesssim 6$.
Previous results implementing this finite-cylinder system on NISQ devices have shown that quantum states with features consistent with Laughlin-type FQH states can be realized in this parameter range~\cite{Rahmani2020, Kirmani2022}. 
An important feature of the truncated Hamiltonian is analytic tractability since its ground state can be written exactly:
\begin{equation} \label{eq:groundstate}
    \ket{\Psi_0} = \mathcal{N} \prod_{j=0}^{N_\phi-1}\left(\hat{1}-t e^{i\phi}\hat{S}_j\right)\ket{100100\dots},
\end{equation}
where $\ket{100100\dots}$ is expressed in the occupation basis of degenerate lowest-Landau-level orbitals, $t=\sqrt{V_{30}/V_{10}}$ and $\phi = 8\pi^2\ell_B^2g^{12}/(g^{11}L_y^2)$ carry dependence on $g^{11}$ and $g^{12}$, the normalization $\mathcal{N}$ depends on $t$ and the {number of electrons $N_e=\nu N_\phi$}, and $\hat{S}_j = \hat{c}^\dagger_{j+1}\hat{c}^\dagger_{j+2}\hat{c}_{j+3}\hat{c}_j$ is an operator that ``squeezes'' neighboring electrons together. 
Using \cref{eq:groundstate} in \cref{eq:viscosity}, the Hall viscosity of the truncated Hamiltonian for the fractional Hall state on a finite cylinder can be computed exactly:
\begin{equation} \label{eq:TTviscosity}
    \eta^A_\mathrm{trunc} = -\frac{64\pi^4\ell_B^2}{\mathcal{A}\,L_y^4}\,t\pdv{}{t}\left(t\pdv{}{t}\ln\mathcal{N}\right),
\end{equation}
which provides an important point of comparison. 
The normalization $\mathcal{N}$ of $\ket{\Psi_0}$ and this expression for the viscosity are derived explicitly in Appendix \ref{app:B}.

\section{Hardware implementation of Hall viscosity}
To estimate the Hall viscosity, we utilize an algorithm that can create the ground states $\ket{\Psi_0}$ for different metrics. 
A linear-depth circuit was found in \cite{Rahmani2020} to create the state [\eqref{eq:groundstate}] for $g_{12}=0$. 
To simplify the circuit, we represent the system in terms of `reduced qubits', each corresponding to a block of 3 consecutive Landau orbitals \cite{Rahmani2020}. 
In this approach, the root pattern used as an initial state is represented as $\ket{100100...}\to\ket{\mathbb{00}...}$ and a squeezed block of states is represented as $\ket{011}\to\ket{\mathbb{1}}$. 
We map each block to a qubit so that we can describe a system of $N_e$ electrons in a space of $N_\phi = 3 N_e$ Landau orbitals using only $N_e$ qubits instead of $N_\phi$. 

\begin{figure}
  \includegraphics[width=\linewidth]{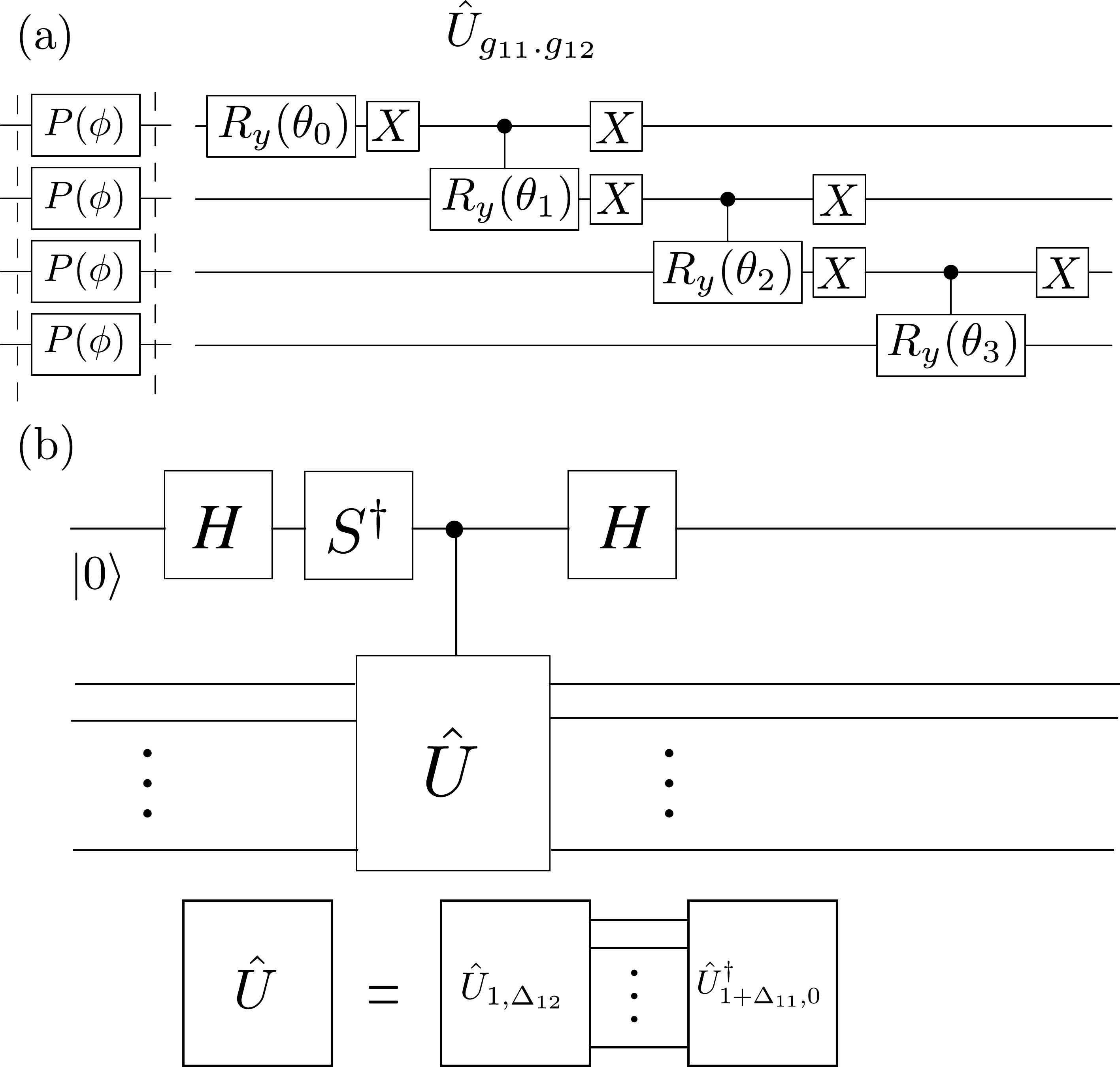}
  \caption{Quantum circuit showing the unitary that creates the ground state with $g^{11}$ and off-diagonal metric $g^{12}$ on a cylinder. Figure (a) show the setup of the Hadamard test for the imaginary part. (b) Shows the decomposition of the Unitary for the first term in Eq.~\eqref{eq:finite diff}. }\label{fig:circ}
\end{figure}

 In the reduced-register representation, the amplitude of a basis state containing $k$ squeezed blocks (i.e., $k$ $\mathbb{1}$ registers) is proportional to $t^k$. 
 The algorithm in Ref.~\cite{Rahmani2020} exploits the fact that two adjacent blocks cannot both be squeezed, since applying a squeezing operator next to an already squeezed block annihilates the state. 
 Accordingly, the procedure begins with the leftmost reduced register, on which a $y$ rotation is applied to create a superposition of $\ket{\mathbb{0}}$ and $\ket{\mathbb{1}}$. 
 A sequence of controlled $y$ rotations is then applied to the subsequent registers, where each rotation is conditioned on the preceding register being in the $|\mathbb{0}\rangle$ state. 
 The corresponding unitary operator is
\begin{equation}\label{eq:unitary_no_phi}
    CR_{y,\mathbb{0}}^{N_e-2,N_e-1}(\theta_{m-1})...CR_{y,\mathbb{0}}^{0,1}(\theta_{1})R_y(\theta_0),
\end{equation}
where $CR_{y,\mathbb{0}}^{i,j}(\theta)$ is a control gate applying a $y$ rotation $R_y(\theta)$ on register $j$ when register $i$ is $\mathbb{0}$. 
In terms of the standard control gate $CR_{y,\mathbb{1}}^{i,j}$ that applies when the control register is $\mathbb{1}$, we have $CR_{y,\mathbb{0}}^{i,j}(\theta)=X_iCR_{y,\mathbb{1}}^{i,j}(\theta)X_i$. 
As was shown in~\cite{Rahmani2020}, the rotation angles follow from the recursion relation $\theta_{n-1}=\arctan\left[-t\cos(\theta_n)\right]$ with the boundary condition $\theta_{N_e-1}=\arctan(-t)$. This unitary creates the state \eqref{eq:groundstate} for $\phi=0$. 
To obtain the ground state for an arbitrary $\phi$, we notice that any term in the superposition with $k$ squeezed, i.e., $\mathbb{1}$, registers must be multiplied by $e^{ik\phi}$. Thus, to transform the state for $\phi=0$ to a state with finite $\phi$, we only need to apply a phaseshift gate $P(\phi)$ with $P(\phi)\ket{\mathbb{0}}=\ket{\mathbb{0}}$ and $P(\phi)\ket{\mathbb{1}}=e^{i\phi}\ket{\mathbb{1}}$ to every register. 
This structure is shown in Fig.~\ref{fig:circ}(a).

According to \cref{eq:viscosity}, we need a quantum circuit that allows us to evaluate $\Im\braket{\pdv{\Psi_0}{g^{12}}}{\pdv{\Psi_0}{g^{11}}}\big|_{g^{11}=1, g^{12}=0}$.
We approximate the above derivatives using finite differences as follows:
\begin{align}
&4 \, \Delta_{11} \, \Delta_{12} \, \braket{\partial_{g^{12}}\Psi_{0}}{ \partial_{g^{11}}\Psi_{0}} \approx \nonumber \\
&\ \ \ \braket{\Psi_0(g^{11}=1, g^{12} = \Delta_{12})}{\Psi_0(g^{11}=1 + \Delta_{11}, g^{12} = 0)} \nonumber \\
&- \braket{\Psi_0(g^{11}=1, g^{12} = -\Delta_{12})}{\Psi_0(g^{11}=1 + \Delta_{11}, g^{12}=0)} \nonumber \\
&- \braket{\Psi_0(g^{11}=1, g^{12} = \Delta_{12})}{\Psi_0(g^{11}=1 - \Delta_{11}, g^{12} = 0)} \nonumber \\
&+ \braket{\Psi_0(g^{11}=1, g^{12} = -\Delta_{12})}{\Psi_0(g^{11}=1 - \Delta_{11}, g^{12}=0)}, \label{eq:finite diff}
\end{align}
with $\Delta_{11} = \Delta_{12}$ chosen small enough to provide a good approximation to the derivative (see Appendix \ref{app:E}).
We can, thus, compute the Hall viscosity on a quantum computer using four distinct circuits, each calculating the imaginary part of the overlaps above~\footnote{Amplitudes are odd in $\Delta_{12}$, i.e.~$\braket{\Psi_0(g^{12}=-\Delta_{12})}{\Psi'_0} = - \braket{\Psi_0(g^{12}=\Delta_{12})}{\Psi'_0}$, so it is possible to estimate the Hall viscosity using only two distinct circuits. However, for each $g^{11}$, we choose to divide our shots across the distinct circuits with $\pm \Delta_{12}$ to average out potential coherent noise in the phase gates.}.

We estimate the imaginary part of each overlap in \cref{eq:finite diff} by applying the Hadamard test, as shown in Fig.~\ref{fig:circ}(b). Writing the matrix elements as
\begin{align}
&\braket{\Psi_0(1, \pm \Delta_{12})}{\Psi_0(1 \pm \Delta_{11}, 0)} \nonumber\\
&= \langle\mathbb{00...0}  | U^\dagger(1, \pm \Delta_{12}) \, U(1 \pm \Delta_{11}, 0)  | \mathbb{00...0} \rangle,
\end{align}
where the unitary $U(g^{11}, g^{12})$ prepares the vacuum state (in the reduced representation) starting from the all-zero state, we first prepare an ancilla qubit in the $|0\rangle - i |1\rangle$ state, then apply $U^\dagger(1, \pm \Delta_{12}) \, U(1 \pm \Delta_{11}, 0)$ to the system qubits controlled on the ancilla, apply a Hadamard gate to the ancilla, and finally estimate $\langle Z \rangle$ on the ancilla qubit.

\begin{figure}
  \includegraphics[width=\linewidth]{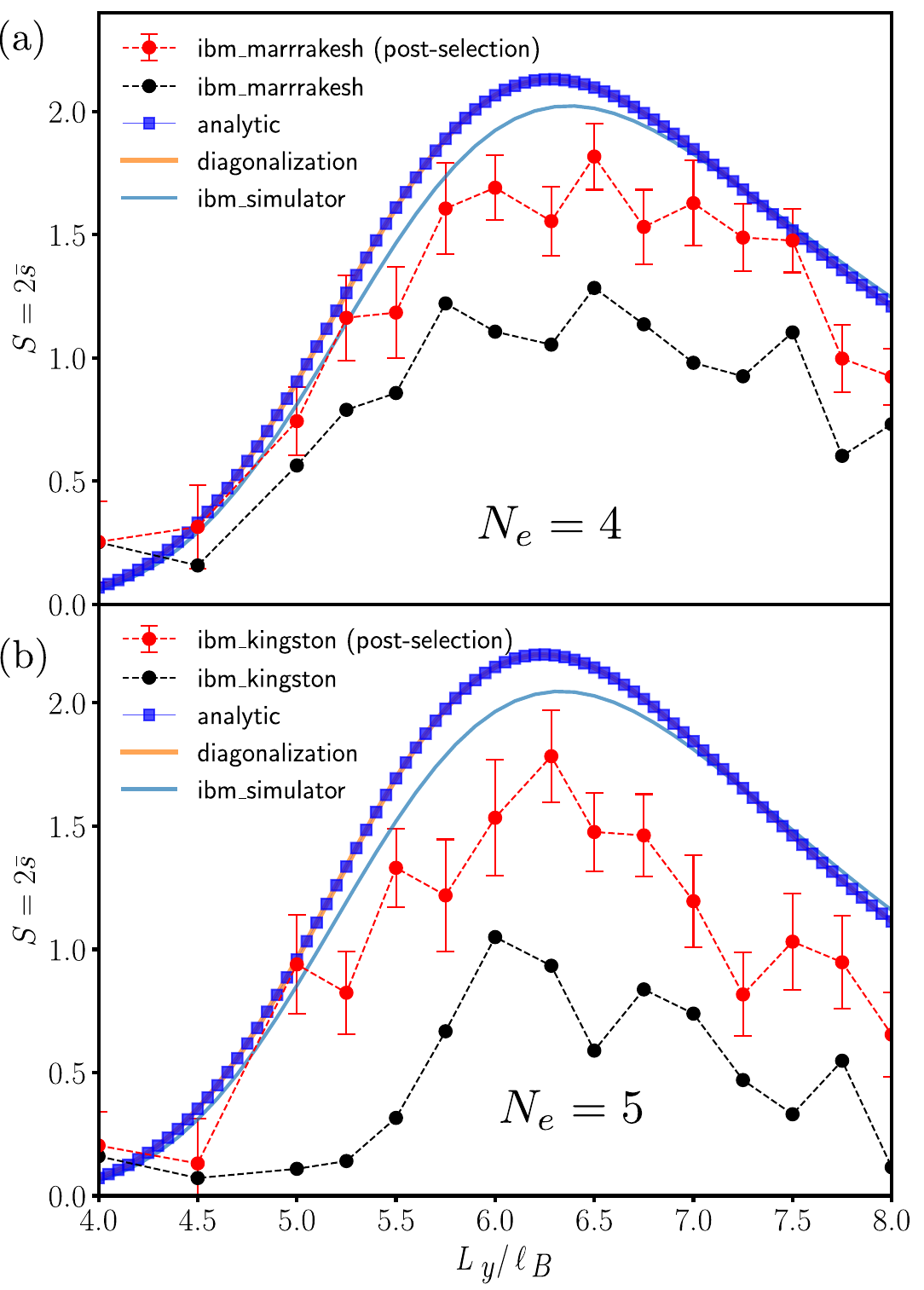}
  \caption{Spin shift $S$ evaluated on a cylinder with circumference $L_y$ for $N_e=4$. 
  The finite difference is taken as $\Delta=0.1$. 
  We show the simulation of the quantum algorithm (blue curve), and the {IBM quantum hardware} results with (red points) and without (black points) post-selection.
  We also plot $S$ obtained from both analytic calculation [green points, proportional to \cref{eq:TTviscosity}] and exact diagonalization of \cref{eq:Htrunc}, demonstrating good agreement between theory and quantum hardware results.}
  \label{fig:IBM}
\end{figure}

We use a modified version of the Hadamard test that reduces the effects of noise. 
In the standard Hadamard test, the $H$ and $S^\dagger$ gates create a state ${1\over \sqrt{2}}\left(|0\rangle-i|1\rangle\right)\otimes |\psi\rangle$ before the application of the controlled unitary. 
After the controlled $U$, the state becomes ${1\over \sqrt{2}}\left(|0\rangle\otimes |\psi\rangle-i|1\rangle\otimes U|\psi\rangle\right)$. 
The last Hadamard finally leads to ${1\over 2}\left[|0\rangle\otimes (I-iU)|\psi\rangle+|1\rangle\otimes (I+iU)|\psi\rangle\right]$.
The expectation value of $Z$ for the ancilla can then give imaginary part of $\expval{U}{\psi}$ because the probability of giving $\pm 1$ ($Z\ket{0}=+1\ket{0}$) is ${1\over 4}\langle \psi|(I\pm iU^\dagger)(I\mp iU)|\psi\rangle$, leading to $\langle Z\rangle={1\over 2}\langle \psi|(iU^\dagger-i U)|\psi\rangle$. 
Our variation is that instead of sampling all states to get $\langle Z\rangle$, we only sample the terms with the working registers having state $\ket{\psi} \ket{\mathbb{00...0}}$, which is easy because this is just a basis state in the computational basis. 
The effect of this selection is equivalent to sampling a projected state ${1\over 2}{\cal N}\left[|0\rangle\otimes \Pi(I-iU)\ket{\psi}+\ket{1}\otimes \Pi(I+iU) \ket{\psi}\right]$, where $\Pi=\dyad{\psi}$ is the projection operator onto $\ket{\psi}$, and the normalization constant $\cal N$ is close to 1 for the specific $U$ operators used in determining a numerical derivative.

In Appendix \ref{app:F}, we show that the normalization constant $\cal N$ contributes to Hall viscosity only at orders higher than $\Delta_{11}\Delta_{12}$ and can therefore be set to 1 to leading order.
This projection then does not change $\langle Z\rangle $ of the ancilla because $U$ is simply replaced with $\Pi U$ leading to $\langle Z\rangle={1\over 2}\ev*{(iU^\dagger\Pi-i \Pi U)}{\psi}=\Im\ev*{U}{\psi}$, where we have used $\Pi\ket{\psi}=\ket{\psi}$.

As discussed above, theoretically calculating $\langle Z\rangle$ for the ancilla gives the same result, with and without projection, or equivalently post-selecting the $\ket{\mathbb{00...0}}$ system state. 
As seen from Fig.~\ref{fig:IBM}, however, in practice, the result without projection suffers more substantially from noise-induced errors. 
Using the state of the system plus ancilla, we can show that the probability of system (excluding the ancilla) being in state $\ket{x}$ is $\frac{1}{2}\big(|\langle x\ket{\mathbb{00...0}}|^2+\bra{x}U\ket{\mathbb{00...0}}\bra{\mathbb{00...0}}U^\dagger\ket{x} \big)$ and the small probability for $\ket{x}$ other than $\ket{\mathbb{00...0}}$ makes them quite fragile to noise.

Using the above methodology, we compute the Hall viscosity as a function of $L_y$ for a system of $N_e = 4$ electrons ($N_\phi = 12$) using (3+1) qubits, including the ancilla. 
We run our calculations on IBM's \textit{ibm\_marrakesh} backend and use the Qiskit transpiler to map the circuits to IBM quantum hardware.

\section{Error-mitigation strategies}
We use the Qiskit Runtime Sampler primitive with dynamical decoupling and Pauli-twirling (randomized compiling) to obtain measured samples for each circuit. 
For each value of $L_2$, we run four circuits to get the Hall viscosity and each circuit is run with 256 random twirls and 1250 shots per twirl for a total of 320000 executions per circuit. 
Due to prohibitive depths of the circuit for NISQ {hardware}, other error mitigation methods like zero-noise extrapolation (ZNE) only degrade the quality of the simulations.

\section{Towards a quantized Hall viscosity}
That our main result (\cref{fig:IBM}) does not reproduce the quantized behavior of $\eta^A$, known for FQH states on a torus, is not a failure of our quantum protocol; the lack of quantization is because the analysis begins from the truncated Hamiltonian \cref{eq:Htrunc}.
Even the exact result [\cref{eq:TTviscosity}] for $\eta^A$ using the analytic ground state of the truncated Hamiltonian is not quantized, demonstrating that the possibility of quantization is lost due to the approximations inherent in the Hamiltonian itself.
The good agreement between {IBM quantum hardware} results and  analytic results in \cref{fig:IBM} demonstrates the success of the quantum protocol to simulate the state $\ket{\Psi_0}$ and to calculate $\eta^A$ with it. 
A secondary obstacle to reproducing the quantization of $\eta^A$ is that quantum computations are limited to the geometry of a finite cylinder, not a torus.
Analytic calculations using the truncated Hamiltonian show a significant dependence on $N_e$ in the finite cylindrical geometry compared to the torus, with the two coinciding in the $N_e\to\infty$ limit---the infinite cylinder and torus results are indistinguishable (see Appendix \ref{app:C}).

In conclusion, we demonstrated the Hall viscosity of Laughlin-type correlated wavefunction on current noisy quantum computers. 
Our results show that currently available quantum computing devices provide a powerful platform demonstrate phenomena that are challenging to execute in conventional experimental settings. 

\section*{Acknowledgments}
\label{sec:acknowledgments}

We thank Tom Iadecola for helpful discussions and suggesting the use of the Hadamard test in our quantum circuit. We acknowledge Roland de Putter from IBM’s Quantum Algorithm Engineering team for his insights and contributions to reducing errors in the quantum hardware calculations. 
This work was supported by National Science Foundation (NSF) Grant No. DMR-2315063 (A.A., P.G. and S.G.) and NSF Grant No. DMR-2315064 (A.R.). 
The work at Los Alamos National Laboratory (LANL) was carried out under the auspices of the U.S. Department of Energy (DOE) National Nuclear Security Administration (NNSA) under Contract No. 89233218CNA000001. 
It was supported by LANL LDRD Program under Project No. 20230049DR  (A.K.), the NNSA Advanced Simulation and Computing Program (J.-X.Z.), Quantum Science Center, a U.S. DOE Office of Science Quantum Information Science Research Center (A.K.). 
It was supported in part by Center of Integrated Nanotechnologies, a DOE Basic Energy Sciences user facility. 
The classical simulations part of the results used the Darwin testbed at LANL, which is funded by the Computational Systems and Software Environments subprogram of LANL Advanced Simulation and Computing program (NNSA/DOE). 
This research used resources provided by the LANL Institutional Computing Program, which is supported by the U.S. Department of Energy National Nuclear Security Administration under Contract No. 89233218CNA000001. 

\section*{Data Availability}
There are no publicly available research data or software supporting this manuscript. 
Requests for further information or data should be sent to the authors.

\appendix

\section{Sheared Torus Metric and Landau-level Wave Functions}
\label{app:A}
We start from a torus with rectangular coordinates $(x',y') \in [0,L_x)\times[0,L_y)$ and metric $g'_{ij}=\mathrm{diag}(1,1)$. 
We assume $L_x \neq L_y$ in general, allowing for the thin torus limit $L_x\gg L_y$ eventually.
We then perform a shearing and scaling coordinate transformation on the torus so that the transformed coordinates $(x,y)\in[0,\tilde{L}_x)\times[0,\tilde{L}_y)$ are related to the rectangular coordinates as
\begin{equation}
    \left(\mqty{x' \\ y'}\right) = \left(\mqty{J_{11} & J_{12} \\ 0 & J_{22}}\right) \left(\mqty{x \\ y}\right).
\end{equation}
The Jacobian of this transformation is $J_{11}J_{22}$ so we take $J_{11} = 1/J_{22}$ to make it area-preserving.
The physical area of the torus is then $\mathcal{A} = L_x L_y = \tilde{L}_x \tilde{L}_y$.
The metric in the transformed coordinates is
\begin{equation}
    g_{ij} = \left(\mqty{1/J_{22}^2 & J_{12}/J_{22} \\ J_{12}/J_{22} & J_{12}^2 + J_{22}^2 }\right),
\end{equation}
with $\mathrm{det}g_{ij} = 1$, and the metric in the dual space is the matrix inverse,
\begin{equation}
    g^{ij} = \left(\mqty{J_{12}^2 + J_{22}^2 & -J_{12}/J_{22} \\ -J_{12}/J_{22} & 1/J_{22}^2}\right).
\end{equation} 

As in Refs.~\cite{avron1995viscosity, Tokatly2009, Pu2020}, we can also parametrize the transformation in terms of complex $\tau = \tau_1 + i \tau_2$; representing rectangular coordinates on the torus as complex numbers, we have $z = x'+i y' = \frac{1}{\sqrt{\tau_2}}(x+\tau y)$, giving $J_{22} = 1/J_{11}= \sqrt{\tau_2}$ and $J_{12} = \tau_1/\sqrt{\tau_2}$. 
In this representation, $\tau_1$ parametrizes a skew transformation and $\sqrt{\tau_2}$ parametrizes an area-preserving scaling of the coordinates.
The metric and its dual in the $\tau$ representation are
\begin{equation}
    g_{ij} = \frac{1}{\tau_2}\left(\mqty{1 & \tau_1 \\ \tau_1 & \abs{\tau}^2}\right) \qquad g^{ij} = \frac{1}{\tau_2}\left(\mqty{\abs{\tau}^2 & -\tau_1 \\ -\tau_1 & 1}\right).
\end{equation}

Writing the Landau Hamiltonian for electrons on this torus pierced by perpendicular magnetic field $B$, the wave function for degenerate states of the lowest Landau level (LLL) in the Landau gauge $\mathbf{A} = (0,B\,x,0)$ is
\begin{multline} \label{eq:SupLLLwf}
    \phi_k(x,y) = \frac{1}{\sqrt{\tilde{L}_y \sqrt{\pi \ell_B^2 g^{11}}}}\sum_{t\in\mathbb{Z}} e^{2\pi i (k+t N_\phi) y/\tilde{L}_y} \\ \times e^{-(1-ig^{12})\frac{\left(x-\tilde{L}_x (k+t N_\phi)/N_\phi\right)^2}{2\ell_B^2 g^{11}}}
\end{multline}
with $k = 0,1,\dots N_\phi-1$, magnetic length $\ell_B = 1/\sqrt{eB}$, and Landau level degeneracy $N_\phi = \mathcal{A}/(2\pi\ell_B^2)$. 
In this gauge choice, the degenerate states parameterized by $k$ are spaced along $x$.

\section{Interaction Hamiltonian}
\label{app:B}
The interaction Hamiltonian for electrons within the LLL is
\begin{equation} \label{eq:eeinteraction}
    H_{ee} = \int\dd\mathbf{R} \int \dd\mathbf{r} \,V(\mathbf{r}) :\hat{n}(\mathbf{R}+\mathbf{r}/2)\hat{n}(\mathbf{R}-\mathbf{r}/2):,
\end{equation}
where $\hat{n}(\mathbf{r}) = \hat{\psi}^\dagger(\mathbf{r})\hat{\psi}(\mathbf{r})$ is the electron density operator in terms of field operators $\hat{\psi}(\mathbf{r})$, and $V(\mathbf{r}) \propto \nabla^2\delta(\mathbf{r})$ is the Trugman-Kivelson pseudopotential for describing fractional quantum Hall states~\cite{Trugman1985}. 
Now, expand the field operators as $\hat{\psi}(\mathbf{r}) = \sum_k \phi_k(\mathbf{r}) \hat{c}_k$, so that $\hat{c}_k$ and $\hat{c}_k^\dagger$ destroy and create electrons in the degenerate states of the LLL.
The calculation can proceed in two ways, giving different looking but equivalent results.
Directly performing the sums with $\tilde{L}_x\gg \ell_B$, we obtain~\cite{Nakamura2012}
\begin{equation} \label{eq:SupHNakamura}
    H_{ee} = \sum_{n=1}^{N_\phi}\sum_{k>\abs{m}} V_{km} \,\hat{c}^\dagger_{n+m} \hat{c}^\dagger_{n+k} \hat{c}_{n+m+k} \hat{c}_n,
\end{equation}
with electron operator indices labeling Landau orbitals defined up to $N_\phi$ and matrix elements
\begin{equation} \label{eq:SupVkm}
    V_{km} \propto (k^2-m^2) e^{-2\pi^2\ell_B^2(k^2+m^2 - 2i g^{12}mk)/(g^{11}\tilde{L}_y^2)},
\end{equation}
as given in \cite{Kirmani2022} for nontrivial but specially uniform metric.

If we, instead, transform the densities and potential in \cref{eq:eeinteraction} to momentum space before expanding the field operators, we more naturally obtain \cref{eq:HYoshioka}, with the metric appearing in the matrix elements \cref{eq:matrixA}. 
This representation is more amenable to direct numerical simulation, while \cref{eq:SupHNakamura} is more easily approximated for further analytic calculations. 

In both forms of the interaction Hamiltonian the index of degenerate LLL states also labels sites in an effective 1d system.
Looking to the wave functions \cref{eq:SupLLLwf}, the orbitals are localized in $x$ at a discrete set of points, $x_k = 2\pi\ell_B^2 k/\tilde{L}_y$ for $k=0,1,\dots N_\phi-1$.
The exponential decay of $V_{km}$ with increasing $k$ and $m$ is, thus, a manifestation of the short-range nature of the Trugman-Kivelson potential. 

\subsection{Truncated Hamiltonian and its Ground State}
For small deformations of the torus $g^{12}\ll 1, g^{11} \approx 1$, the thin-torus limit $L_y\ll L_x$ also gives $\tilde{L}_y\ll \tilde{L}_x$. 
The matrix elements $V_{km}$ become exponentially small for large values of $k$ and $m$ if $2\pi^2 \ell_B^2/(g^{11}\tilde{L}_y^2)\gtrsim 1$, equivalent to the statement $\tilde{L}_x \gtrsim g^{11}N_\phi \tilde{L}_y/\pi$ which gives a lower bound to what is meant by the thin-torus limit.
Taking filling fraction $\nu=1/3$ so that the number of electrons is $N_e = N_\Phi/3$, to a good approximation we may discard terms in \cref{eq:SupHNakamura} with large $k$ and $m$ and obtain the \emph{truncated} Hamiltonian~\cite{Nakamura2012},
\begin{multline} \label{eq:SupHtrunc}
    H_\mathrm{trunc} = \sum_{j=0}^{N_\phi-1} V_{10} \hat{n}_j\hat{n}_{j+1} + V_{20}\hat{n}_j\hat{n}_{j+2} \\  + V_{30}\hat{n}_j\hat{n}_{j+3}+V_{21}\hat{c}^\dagger_{j+1} \hat{c}^\dagger_{j+2} \hat{c}_{j+3} \hat{c}_j + \mathrm{h.c. },
\end{multline}
which has exact ground state
\begin{equation} \label{eq:Suppsi0}
    \ket{\Psi_0} = \mathcal{N} \prod_{j=0}^{N_\phi-1}\left(\hat{1}-t e^{i\phi}\hat{S}_j\right)\ket{\dots100100\dots}.
\end{equation}
The reference state $\ket{\dots100100\dots}$ is written in the occupation basis of the degenerate lowest Landau level states, $\mathcal{N}$ is a normalization factor to be determined, $t=\sqrt{V_{30}/V_{10}} = \exp\left[-8\pi^2\ell_B^2/(g^{11}\tilde{L}_y^2)\right]$, $\phi = 8\pi^2\ell_B^2g_{12}/(g^{11}\tilde{L}_y^2)$, and $\hat{S}_j = \hat{c}^\dagger_{j+1}\hat{c}^\dagger_{j+2}\hat{c}_{j+3}\hat{c}_j$ is an operator that ``squeezes'' neighboring electrons together.

Since $\ket{\dots100100\dots}$ has electrons only at positions $j=3k$ for $k=0,1,2,\dots N_e-1$, only $\hat{S}_{3k}\equiv \hat{\mathbb{S}}_k$ have nontrivial action.
Furthermore, acting with $\hat{\mathbb{S}}_k$ forbids nontrivial action of $\hat{\mathbb{S}}_k\pm1$, where $k$ is defined mod $N_e$ on the torus.
Defining effective ``bits'' by $\ket{1_{3k}0_{3k+1}0_{3k+2}} \equiv\ket{\mathbb{0}_k}$, $\ket{0_{3k}0_{3k+1}0_{3k+2}} \equiv \ket{\mathbb{0}_k}$, and $\ket{0_{3k}1_{3k+1}1_{3k+2}} \equiv \ket{\mathbb{1}_k}$, this constraint can be rephrased as forbidding ``bitstrings'' with adjacent $\mathbb{1}$s, where $k=0$ and $N_e-1$ are adjacent. 
The product generating the ground state is thus a sum over all possible states obeying this constraint, with each term acquiring a factor of $-t e^{i\phi}$ every time a $\hat{\mathbb{S}}$ operator acts nontrivially flipping a $\mathbb{0}$ to a $\mathbb{1}$.
We write
\begin{equation}
    \ket{\Psi_0} = \mathcal{N} \sum_{\mathbb{B}\in \mathcal{H}(N_e)} (-\alpha)^{\hat{n}_\mathbb{1}}\ket{\mathbb{B}} \equiv \mathcal{N}\vert\tilde{\Psi}_0\rangle,
\end{equation}
where $\alpha = te^{i\phi}$, $\mathcal{H}(N_e)$ is the Hilbert space of states obeying the local constraint, each labeled by the corresponding bitstring $\mathbb{B}$, $\vert\tilde{\Psi}_0\rangle$ is the unnormalized weighted sum over these states, and $\hat{n}_\mathbb{1}$ counts the number of $\mathbb{1}$s in each $\mathbb{B}$.
The dimension of $\mathcal{H}(N_e)$ obeys the Fibonacci recurrence relation, but with different base cases: $\mathrm{dim}[\mathcal{H}(N+2)] = \mathrm{dim}[\mathcal{H}(N+1)] + \mathrm{dim}[\mathcal{H}(N)]$ with $\mathrm{dim}[\mathcal{H}(2)] = 3$ and $\mathrm{dim}[\mathcal{H}(3)] = 4$.

The normalization $\mathcal{N}$ is determined through
\begin{equation}
    \bra{\Psi_0}\ket{\Psi_0} = 1 = \abs{\mathcal{N}}^2 \sum_{n=0}^{\lfloor{N_e/2}\rfloor}C^{N_e}_n\,t^{2n},
\end{equation}
so that that $\bra{\tilde{\Psi}_0}\ket{\tilde{\Psi}_0}=\mathcal{N}^{-2}$, where $C^{N}_n$ is the number of states in $\mathcal{H}(N)$ with exactly $n$ $\mathbb{1}$s in their bitstring representation. 
A combinatoric argument incorporating periodic boundary conditions gives
\begin{equation}
    C^N_n = \frac{N}{N-n} \begin{pmatrix} N-n \\ n \end{pmatrix},
\end{equation}
in terms of a binomial coefficient, and using this inside the sum then gives
\begin{widetext}
\begin{equation}\label{eq:Supnorm}
    \mathcal{N}(N_e,t) = \begin{cases}
        N_e^{-1/2} \left[\gamma\left(\frac{N_e+1}{2},t\right)-t^2\gamma\left(\frac{N_e-1}{2},t\right) + (N_e-1)\left(\frac{1+\sqrt{1+4t^2}}{2}\right)^{N_e}\right]^{-\frac{1}{2}}, & N_e\,\,\mathrm{odd} \\
        \gamma\left(\frac{N_e}{2},t\right)^{-\frac{1}{2}}, & N_e \,\,\mathrm{even} 
    \end{cases}
\end{equation}
in terms of
\begin{equation}
    \gamma(Q,t) = \left(\frac{1+2t^2+\sqrt{1+4t^2}}{2}\right)^Q+\left(\frac{1+2t^2-\sqrt{1+4t^2}}{2}\right)^Q.
\end{equation}
Through its dependence on $t$, the normalization factor $\mathcal{N}$ is, thus, a function of the metric component $g^{11}$ but not $g^{12}$.
\end{widetext}

\subsection{Finite cylinder geometry}
If instead of a torus we consider a finite cylindrical geometry by simply not imposing periodic boundary conditions along the $x$ direction, then, ignoring edge states, the system obeys a similar truncated Hamiltonian \cref{eq:SupHtrunc} without the periodic condition on the fermionic indices.
The ground state of this Hamiltonian is identical to \cref{eq:Suppsi0}, but again without periodicity mod $N_\phi$. 
As a result, the final effective ``bit'' $\ket{\mathbb{0}_{N_e-1}}= \ket{1_{N_\phi-3} 0_{N_\phi-2} 0_{N_\phi-1}}$ can never be put into the state $\ket{\mathbb{1}_{N_e-1}}$. 
This lack of periodic boundary conditions changes the allowed states for each $N_e$, but similar combinatoric analysis gives the normalization
\begin{equation} \label{eq:Supcylnorm}
    \mathcal{N}_\mathrm{c}(N_e,t) = \begin{cases}
        \gamma_\mathrm{c}\left(\frac{N_e+1}{2},t\right)^{-\frac{1}{2}}, & N_e\,\,\mathrm{odd} \\
        \left[\gamma_\mathrm{c}\left(\frac{N_e+2}{2},t\right) - t^2\gamma_\mathrm{c}\left(\frac{N_e}{2},t\right)\right]^{-\frac{1}{2}}, & N_e\,\,\mathrm{even}
    \end{cases}
\end{equation}
with
\begin{equation}\begin{split}
    & \gamma_\mathrm{c}(Q,t) =  \frac{1}{\sqrt{1+4t^2}}\\ & \left[\left(\frac{1+2t^2+\sqrt{1+4t^2}}{2}\right)^Q-\left(\frac{1+2t^2-\sqrt{1+4t^2}}{2}\right)^Q\right].
\end{split} \end{equation}

\section{Hall Viscosity}
\label{app:C}
In Refs.~\cite{avron1995viscosity,Pu2020}, the relationship between the Hall viscosity and the Berry curvature in the 2d space of $\tau_1$ and $\tau_2$ is given as
\begin{equation}
    \eta^A = -\frac{\tau_2^2}{\mathcal{A}} \mathcal{F}_{\tau_1\tau_2} = \frac{2\tau_2^2}{\mathcal{A}}\Im\braket{\pdv{\Psi_0}{\tau_1}}{\pdv{\Psi_0}{\tau_1}}.
\end{equation}
Using the explicit expressions for the independent components of the inverse metric $g^{11}=\abs{\tau}^2/\tau_2$ and $g^{12}=-\tau_1/\tau_2$, we find
\begin{equation} \begin{split}
    \Im\braket{\pdv{\Psi_0}{\tau_1}}{\pdv{\Psi_0}{\tau_2}} & = \abs{\frac{\partial(g^{11},g^{12})}{\partial(\tau_1,\tau_2)}} \Im\braket{\pdv{\Psi_0}{g^{11}}}{\pdv{\Psi_0}{g^{12}}} \\ & = \frac{\abs{\tau}^2}{\tau_2^3} \Im\braket{\pdv{\Psi_0}{g^{11}}}{\pdv{\Psi_0}{g^{12}}} \\ & = \frac{g^{11}}{\tau_2^2} \Im\braket{\pdv{\Psi_0}{g^{11}}}{\pdv{\Psi_0}{g^{12}}},\end{split}\label{eq:Suptautog} 
\end{equation}
so that the Hall viscosity in terms of $g^{11},g^{12}$ is
\begin{equation} \label{eq:Supviscosityg}
    \eta^A =\frac{2g^{11}}{\mathcal{A}} \Im\braket{\pdv{\Psi_0}{g^{11}}}{\pdv{\Psi_0}{g^{12}}} = -\frac{g^{11}}{\mathcal{A}}\mathcal{F}_{g^{11}g^{12}},
\end{equation}
evaluated at $g^{11}=1, g^{12}=0$.

\subsection{Truncated ground state result}

Using the form of the truncated Hamiltonian ground state and its normalization \cref{eq:Supnorm}, the derivatives are
\begin{equation}\begin{split}
    \ket{\partial_a\Psi_0} = (&\partial_a\mathcal{N})\ket{\tilde{\Psi}_0} + \mathcal{N}\partial_a\ket{\tilde{\Psi}_0} \\  = & (\partial_a\ln\mathcal{N})\ket{\Psi_0} \\ & + \mathcal{N}(\partial_a\ln\alpha(g)) \sum_{\mathbb{B}} \hat{n}_\mathbb{1} (-\alpha(g))^{\hat{n}_\mathbb{1}}\ket{\mathbb{B}},\end{split}
\end{equation}
where $\partial_a\equiv \pdv{}{g^{1a}}$. 
Substituting this into the expression for the curvature, we obtain two inner products which we can calculate:
\begin{gather}
    \mathcal{N}\bra{\Psi_0}\sum_\mathbb{B}\hat{n}_\mathbb{1}(-\alpha)^{\hat{n}_\mathbb{1}}\ket{\mathbb{B}} = -t\pdv{\ln\mathcal{N}}{t} \\
    \mathcal{N}^2\sum_{\mathbb{B},\mathbb{B}'}\bra{\mathbb{B}} \hat{n}_\mathbb{1}^2\,t^{2\hat{n}_\mathbb{1}}\ket{\mathbb{B}'} = \frac{1}{4}\mathcal{N}^2t\pdv{}{t}\left(t\pdv{}{t}\frac{1}{\mathcal{N}^2}\right)
\end{gather}
so that the full curvature tensor is
\begin{equation} \begin{split}
    \mathcal{F}_{ab} = & -2t\pdv{\ln\mathcal{N}}{t}(\partial_a\ln\mathcal{N})\Im\left[\partial_b\ln\alpha(g) +\partial_a\ln\alpha(g)^\ast\right] + \\ & \frac{\mathcal{N}^2}{2}t\pdv{}{t}\left(t\pdv{}{t}\frac{1}{\mathcal{N}^2}\right)\Im\left[(\partial_a\ln \alpha(g)^\ast) (\partial_b\ln \alpha(g))\right], \end{split}
\end{equation}
now expressed entirely in terms of the normalization factor $\mathcal{N}$ and the complex coefficient $\alpha(g)$ appearing in the ground state. 
We have used the fact that $\mathcal{N}$ is real to simplify. 
We can see $\mathcal{F}_{11}=\mathcal{F}_{22}=0$ generically.

Using $\alpha=t e^{i\phi}$ and the explicit form of $\phi$, then the Hall viscosity of the truncated Hamiltonian ground state is
\begin{equation} \label{eq:SupViscosity}
    \eta^A_\mathrm{trunc} = - \frac{64\pi^4\ell_B^4}{\mathcal{A}\, L_y^4} \,t\pdv{}{t}\left(t\pdv{}{t}\ln\mathcal{N}\right),
\end{equation} 
with $t=3e^{-8\pi^2\ell_B^2/L_y^2}$ and $\mathcal{N}$ as in \cref{eq:Supnorm} for the torus or \cref{eq:Supcylnorm} for the finite cylinder.
When $t\ll 1$ then $\partial\ln\mathcal{N}/\partial t \approx -t N_e$ for the torus and $\partial\ln\mathcal{N}/\partial t \approx -t (N_e-1)$ for the finite cylinder, so the derivatives can be evaluated simply, giving
\begin{equation} \label{eq:Supapprox}
    \eta^A_\mathrm{trunc} \approx \begin{cases} 1152\pi^4 \frac{N_e \ell_B^4}{\mathcal{A}\,L_y^4}\exp\left(-\frac{8\pi^2\ell_B^2}{L_y^2}\right), & \mathrm{torus}\\
    1152\pi^4 \frac{(N_e-1) \ell_B^4}{\mathcal{A}\,L_y^4}\exp\left(-\frac{8\pi^2\ell_B^2}{L_y^2}\right), & \mathrm{cylinder}
    \end{cases}
\end{equation} 
Note that $t\lesssim 0.1$ for $L_y/\ell_B \lesssim5$, so this approximation applies only for small values of $L_y/\ell_B$.
For larger values, we cannot approximate the derivative in this way and must keep the full expression [\cref{eq:SupViscosity}], which is simple to evaluate numerically. 

\begin{figure}
    \centering
    \includegraphics[width=\columnwidth]{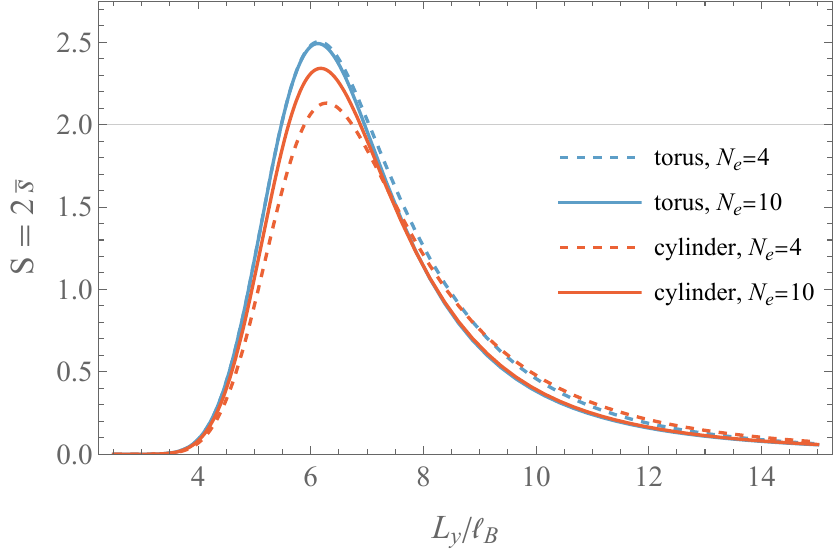}
    \caption{The exact value of the spin shift $S$ computed from the truncated Hamiltonian [\cref{eq:SupHtrunc}] for the torus (blue) and finite cylinder (red) geometries with $N_e=4$ (dashed) and $N_e=10$ (solid).}
    \label{fig:analyticS}
\end{figure}

From Ref.~\cite{read2011hall}, we can relate the Hall viscosity to the mean orbital spin per particle $\bar{s}$ through $\eta^A = \bar{s}N_e/(2\mathcal{A})$, so for the case we are considering,
\begin{equation} \label{eq:Supsbar}
    \bar{s} = -\frac{128\pi^4\ell_B^4}{N_e\,L_y^4}\,t\pdv{}{t}\left(t\pdv{}{t}\ln\mathcal{N}\right) \approx 2304 \pi^4 \frac{\ell_B^4}{L_y^4}e^{-16\pi^2\ell_B^2/L_y^2}, 
\end{equation}
with the approximate result applying for small $L_y/\ell_B$ on the torus.
The spin shift $S$ is related to this as $S = 2\bar{s}$; the exact expression for $S$ from the truncated Hamiltonian is plotted in \cref{fig:analyticS} for both the torus and finite cylinder for $N_e=4$ and $N_e=10$.
The torus has only weak dependence on $N_e$, while the finite cylinder has more dramatic dependence and approaches the torus result as $N_e$ increases.

\section{Validating $\eta^A$: $\tau$ vs $g$ dependence}
\label{app:D}
Here we demonstrate the equivalence in \cref{eq:Suptautog} by comparison to known analytic results. 

\subsection{LLL single particle Berry curvature}
We evaluate the Berry curvature of a single orbital in the lowest Landau level (LLL) using the metric parametrization and compare to the known result in the $\tau$ parametrization~\cite{Levay1995, Pu2020},
\begin{equation}
    \mathcal{F}^{(1P)}_{\tau_1\tau_2} = -\frac{1}{2\tau_2^2}\left(n+\frac{1}{2}\right),
\end{equation} 
for a single particle in the $n$th Landau level. 
Using the LLL wave function [\cref{eq:SupLLLwf}], we have
\begin{equation}
    \mathcal{F}^{(1P)}_{g^{11}g^{12}} = \int \dd \mathbf{r} \,\,\pdv{\phi_k(\mathbf{r})^\ast}{g^{11}} \pdv{\phi_k(\mathbf{r})}{g^{12}} = -\frac{1}{4g^{11}},
\end{equation}
so that
\begin{equation}
    \eval{\mathcal{F}^{(1P)}_{\tau_1\tau_2}}_{n=0} = \frac{g^{11}}{\tau_2^2}\mathcal{F}^{(1P)}_{g^{11}g^{12}},
\end{equation}
agreeing with \cref{eq:Suptautog}. 

\subsection{Comparison to M. Fremling, J. Phys. A 50, 015201 (2016)~\cite{Fremling2016}}

In Ref.~\cite{Fremling2016} $\bar{s}$ is evaluated with an approximation to the state $\ket{\Psi_0}$ parameterized by the modular parameter $\tau$,
\begin{equation}
    \ket{\Psi_0}\approx\ket{\tau} = N(\tau)\left(\ket{TT} + a(\tau)\sum_{i=1}^{N_e}\ket{i,i+1}\right),
\end{equation}
where $\ket{TT} = \ket{\dots 100100\dots}$, $\ket{i,i+1} = \ket{\dots 011000\dots}$ at position $i$, $a(\tau) = 3\exp\left(i\pi\tau \frac{4}{3N_e}\right)$, and $N(\tau)$ is the normalization.  
With this state appropriately normalized, ignoring a constant contribution $1/2$, Ref.~\cite{Fremling2016} finds 
\begin{equation}
    \bar{s} = 2304\pi^4 \frac{e^{16\pi^2/L_y^2}}{L_y^4\left(e^{16\pi^2/L_y^2} + 9 N_e\right)^2},
\end{equation}
where $\ell_B$ has been set to 1.
When making the same approximation to \cref{eq:Suppsi0}, i.e., keeping only the term linear in $t$, and evaluating $\bar{s}$ in terms of derivatives with respect to components of $g$, then we obtain the same result. 
Furthermore, comparing to our approximate result for small $L_y/\ell_B$ in \cref{eq:Supsbar}, we find agreement in the (artificial) $N_e=0$ limit of the result of Ref.~\cite{Fremling2016}, which is noted there to be an idealized upper bound on $\bar{s}$. 

\section{Finite Difference Approximation of $S$}
\label{app:E}
\begin{figure}[!hb]
  \includegraphics[width=\columnwidth]{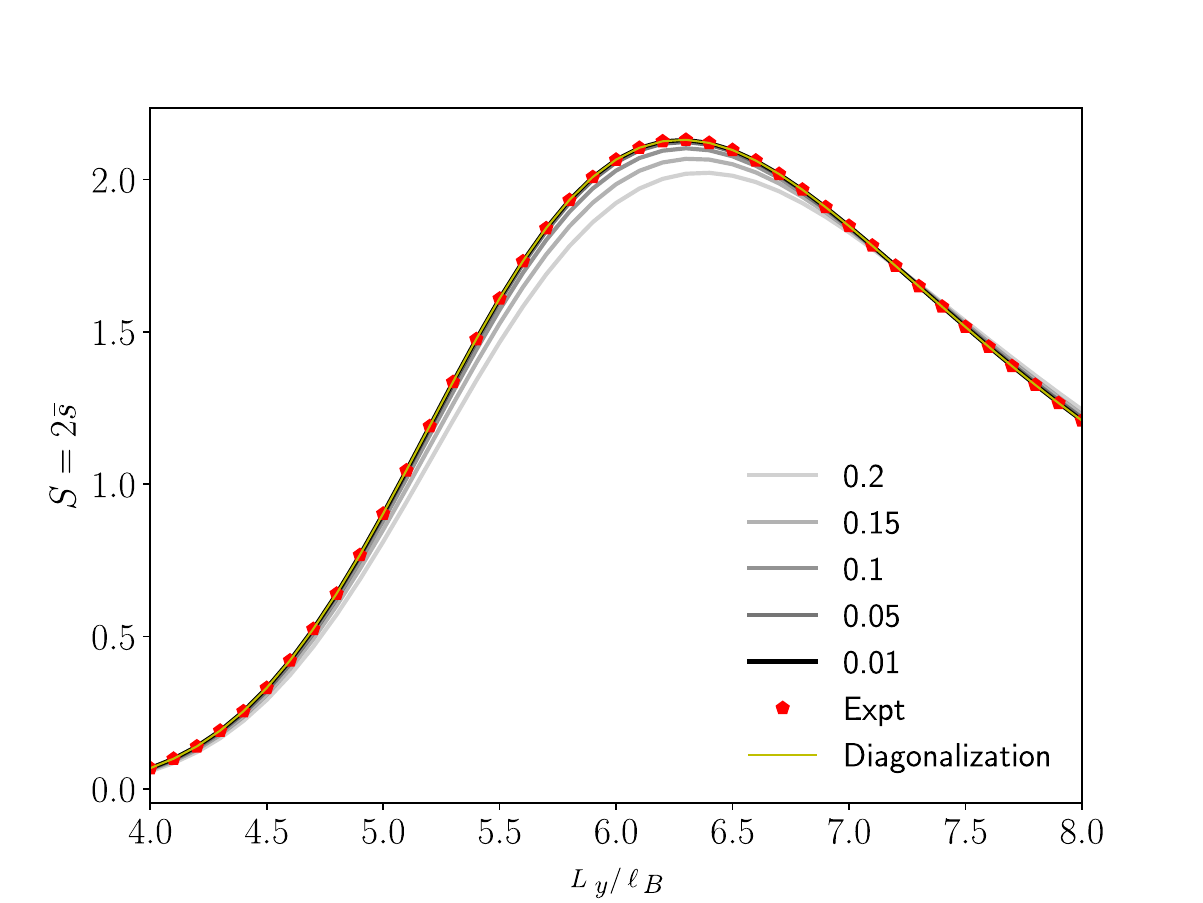}
  \caption{$S$ for various $\Delta=\Delta_{11}=\Delta_{12}$ against $L_y$ obtained from the quantum circuit of the main text on noiseless simulator. The extrapolation is only performed on $\Delta_{11}=\Delta_{12}$ values of 0.2, 0.15, 0.1, and 0.05. Exact diagonalization results agree with the extrapolated results. }
  \label{fig:extrp}
\end{figure}

When implementing \cref{eq:Supviscosityg} on a quantum device, the derivatives---with respect to the metric components---are approximated as finite differences parameterized by the small parameters $\Delta_{11}$, $\Delta_{12}$.
In \cref{fig:extrp}, we show how the value of the spin shift $S$ depends on these parameters.

\section{Justification of post-selection}
\label{app:F}
The quantum state prepared by the Hadamard-test circuit is
\begin{equation}
\ket{\Psi}
=
\frac{1}{2}
\bigg(
\ket{0}_A \otimes (I-i\hat U)\ket{\Psi_0}
+
\ket{1}_A \otimes (I+i\hat U)\ket{\Psi_0}
\bigg),
\end{equation}
where $\ket{\Psi_0}=\ket{00\cdots 0}$ denotes the reference (root) state of the system and $|0,1\rangle_A$ are ancilla states.
Defining
\begin{equation}
\ket{\alpha}=(I-i\hat U)\ket{\Psi_0},
\qquad
\ket{\beta}=(I+i\hat U)\ket{\Psi_0},
\end{equation}
the joint ancilla-system density matrix is
\begin{multline}
\rho_{AS}
=
\ket{\Psi}\bra{\Psi}
=
\frac{1}{4}
\bigg(
\ket{0}\bra{0}\otimes\ket{\alpha}\bra{\alpha}
+
\ket{1}\bra{1}\otimes\ket{\beta}\bra{\beta} \\
+
\ket{0}\bra{1}\otimes\ket{\alpha}\bra{\beta}
+
\ket{1}\bra{0}\otimes\ket{\beta}\bra{\alpha}
\bigg).
\end{multline}
We have dropped the $A$ subscript for brevity.
The ancilla $Z$ expectation value is obtained from the operator $Z_A\otimes I$. Using $\mathrm{Tr}[\rho_{AS}]=1$, we find
\begin{align}
\langle Z_A\rangle
&=
\mathrm{Tr}\!\left[(Z_A\otimes I)\rho_{AS}\right] \nonumber\\
&=
\mathrm{Tr}\!\left[\ket{0}\bra{0}\otimes\rho_{AS}\right]
-
\mathrm{Tr}\!\left[\ket{1}\bra{1}\otimes\rho_{AS}\right] \nonumber\\
&=
\frac{1}{2}\sum_x
\big(
|\langle x|\alpha\rangle|^2
-
|\langle x|\beta\rangle|^2
\big) \nonumber\\
&=
\frac{1}{2}
\big(
\langle\alpha|\alpha\rangle
-
\langle\beta|\beta\rangle
\big)
=
\operatorname{Im}\langle\Psi_0|\hat U|\Psi_0\rangle.
\end{align}

We now consider post-selection onto the system state $\ket{\Psi_0}$, implemented by the projector
\begin{equation}
\Pi_S=\ket{\Psi_0}\bra{\Psi_0}.
\end{equation}
The post-selected (unnormalized) state is
\begin{align}
\ket{\Psi'}
 = &
(I\otimes\Pi_S)\ket{\Psi} \nonumber \\
 = & \frac{1}{2}
\bigg(
\ket{0}_A \otimes \ket{\Psi_0}\bra{\Psi_0}(I-i\hat U)\ket{\Psi_0}  \\
 & +
\ket{1}_A \otimes \ket{\Psi_0}\bra{\Psi_0}(I+i\hat U)\ket{\Psi_0}
\bigg).
\end{align}

Defining
\begin{equation}
z=\langle\Psi_0|\hat U|\Psi_0\rangle,
\end{equation}
this state can be written as
\begin{equation}
\ket{\Psi'}
=
\frac{1}{2}
\bigg(
(1-iz)\ket{0}_A
+
(1+iz)\ket{1}_A
\bigg)
\otimes
\ket{\Psi_0}.
\end{equation}

The corresponding (unnormalized) density matrix is, therefore,
\begin{equation}
\rho_{AS}^{\mathrm{proj}}
=
\frac{1}{4}
\begin{pmatrix}
|1-iz|^2 & (1-iz)(1+iz)^* \\
(1+iz)(1-iz)^* & |1+iz|^2
\end{pmatrix}
\otimes
\ket{\Psi_0}\bra{\Psi_0}.
\end{equation}

The trace of the projected density matrix can be used to normalize the expectation values.
\begin{equation}
\mathrm{Tr}\!\left[\rho_{AS}^{\mathrm{proj}}\right]
=
\frac{1}{4}
\left(
|1-iz|^2+|1+iz|^2
\right).
\end{equation}

Finally, the ancilla $Z$ expectation value conditioned on post-selection is
\begin{align}
\langle Z_A\rangle_{\mathrm{proj}}
= &
\frac{
\mathrm{Tr}\!\left[(Z_A\otimes I)\rho_{AS}^{\mathrm{proj}}\right]
}{
\mathrm{Tr}\!\left[\rho_{AS}^{\mathrm{proj}}\right]
} \nonumber
=
\frac{|1-iz|^2-|1+iz|^2}{|1-iz|^2+|1+iz|^2}
\\ = &
\frac{2\,\operatorname{Im}(z)}{1+|z|^2}.
\end{align}

The overlap we need to compute has two derivatives. Therefore, we need to expand each of them to first order. The infinitesimal variables are independent and we call them $\delta_1$ and $\delta _2$. Therefore, we need to find the following expression to order $\delta_1\delta_2$:
\begin{widetext}\[
\langle\mathbb{00...0}|\hat U(\delta_1,\delta_2)|\mathbb{00...0}\rangle-\langle\mathbb{00...0}|\hat U(-\delta_1,\delta_2)|\mathbb{00...0}\rangle-\langle\mathbb{00...0}|\hat U(\delta_1,-\delta_2)|\mathbb{00...0}\rangle+\langle\mathbb{00...0}|\hat U(-\delta_1,-\delta_2)|\mathbb{00...0}\rangle,
\]\end{widetext}

The post-selection procedure is justified by the structure of the unitary
\[\hat U(\delta_1,\delta_2)=U_0^
\dagger(1+\delta_1)e^{-iG\delta_2}U_0(1),\]
where $U_{0}$ is defined in \cref{eq:eq:unitary_no_phi} of the main text and is real, as it is composed entirely of controlled $R_y$ gates. 
Similarly, $e^{-iG\delta}$ is the tensor product of all $P(\phi)$ gates shown in Fig. 2(a) of the main text. Due to the structure of the $P(\phi)$ gates, $G$ is a real Hermitian matrix.

We reiterate that both unitary $U_0$ and Hermitian $G$ are {\it real} operators in the computational basis.
Expanding $\hat U$ gives 
\begin{multline*}
    \hat U(\delta_1,\delta_2)\approx(\mathbb{1}+(\partial_g U_0^\dagger)\delta_1)(\mathbb{1}-i\delta_2 G)U_0 \\
    =\mathbb{1}+\delta_1(\partial_g U_0^\dagger) U_0-i\delta_2GU_0-i\delta_1\delta_2(\partial_g U_0^\dagger) GU_0
\end{multline*}
We now argue that $\langle (\partial_g U_0^\dagger) U_0\rangle=0$. 
Clearly, $\langle U_0^\dagger U_0\rangle=1$ and taking a derivative gives
\begin{multline*}
0=\langle (\partial_gU_0^\dagger) U_0\rangle+\langle U_0^\dagger (\partial_g U_0)\rangle\\
=\langle (\partial_gU_0^\dagger) U_0\rangle+\left[\langle (\partial_gU_0^\dagger) U_0\rangle\right]^*\\
=2{\rm Re}\left[\langle (\partial_gU_0^\dagger) U_0\rangle\right].
\end{multline*}
Since $U_0$ is a real matrix in the computational basis, we conclude that when the expectation value is taken with respect to the real state $|\mathbb{00...0}\rangle$, we must have $\langle (\partial_g U_0^\dagger) U_0\rangle={\rm Re}\left[\langle (\partial_g U_0^\dagger)U_0\rangle\right]=0$. 
We then write
\begin{widetext}
\[ z(\delta_1,\delta_2)\equiv \langle\mathbb{00...0}|\hat U(\delta_1,\delta_2)|\mathbb{00...0}\rangle\approx 1-i\delta_2\langle \mathbb{00...0}| GU_0|\mathbb{00...0}\rangle-i\delta_1\delta_2\langle \mathbb{00...0}|(\partial_g U_0^\dagger) GU_0|\mathbb{00...0}\rangle \equiv 1-i\delta_2b-i\delta_1\delta_2c. \]
\end{widetext} 
We note that in the combination of interest, the first-order terms cancel out. Namely,
\[
z(\delta_1,\delta_2)-z(-\delta_1,\delta_2)-z(\delta_1,-\delta_2)+z(-\delta_1,-\delta_2)=-4i\delta_1\delta_2c
\]
Let us now consider the result of the Hadamard-test ancilla measurement with post-selection:
\begin{multline*}
{2{\rm Im}\left[z(\delta_1, \delta_2)\right]\over 1+|z(\delta_1, \delta_2)|^2}={-(\delta_2b+\delta_1\delta_2 c)\over 1+\delta_2^2(b+c\delta_1)^2/2} \\
\approx -(\delta_2b+\delta_1\delta_2 c)(1-\delta_2^2(b+c\delta_1)^2/2+...).
\end{multline*}
Any correction from the denominator will be order $\delta^3$ or higher, so indeed the normalization of the projected state is unimportant.

\section{Towards quantized viscosity}
\label{app:G}
As seen from the main text, that quantized value of Hall viscosity is obtained only in the limit $L_y>10\ell_B$ for $N\geq6$. 
While, the ground state [Eq.~\eqref{eq:Suppsi0}] can be prepared in the thin torus limit through a shallow depth circuit for NISQ hardware in our work, preparing the ground state on a torus of arbitrary circumference is beyond the capability of NISQ devices. In what follows, we will denote the ground state of Eq.~\eqref{eq:Suppsi0} as $\ket{\Psi_{TT}}$, the starting TT Hamiltonian as $H_{TT}$ and Hamiltonian for torus of arbitrary thickness given as $H(L_y,g)$~\cite{Yoshioka1983}:
\begin{equation}\label{eq:Sup_HYoshioka}
    H(L_y,g)=\sum_{\{j_i\}=0}^{N_{\phi}-1}A_{j_1,j_2,j_3,j_4}(g)\,\hat{c}^\dagger_{j_1} \hat{c}^\dagger_{j_2} \hat{c}_{j_3} \hat{c}_{j_4}.
\end{equation} 
with the effective interaction potentials
\begin{equation}\begin{split}
 \label{eq:Sup_matrixA}
    A_{j_1j_2j_3j_4}(g)= & \frac{\delta'_{j_1+j_2,j_3+j_4}}{2\mathcal{A}} 
    \times \\ & \sum_{s,t\in \mathbb{Z}}\delta'_{j_1-j_4,t}\left(g^{ij}q_i q_j \right)e^{-\frac{1}{2} \left(g^{ij}q_i q_j\right)}e^{2\pi is\frac{j_1-j_3}{N_\phi}}.
\end{split}\end{equation}
We verify that the ground state to Eq.~\eqref{eq:Sup_HYoshioka} for $L_y\geq10\ell_B$ ($g_{11}=1$ and $g_{12}=0$) can be prepared with the following Schrodinger equation for $s=t/T$:
\begin{multline}\label{eq:Sup_AdiabaticSE}
    -ih\frac{\partial}{T\partial s}\ket{\Psi(s,g)} =\bigg[(1-s)H_{TT}+sH(g)\bigg]\ket{\Psi(s)} \\
    =H(s,g)\ket{\Psi(s)} 
\end{multline}
This is solved by the time-dependent unitary
\begin{equation}\label{eq:Sup_Unitarys}
 U(s,g)=\mathrm{\tau} e^{-i\frac{T}{\hbar}\int_0^{s}H(s',g)ds'}   
\end{equation}
For $T>>1$, we get the ground state of Eq.~\eqref{eq:Sup_HYoshioka} for desired $L_y$ when $s=1$. The unitary in Eq.~\ref{eq:Sup_Unitarys} can be implemented at discrete times $t=t_1,....t_N$ and $dt=t_2-t_1$.

\begin{equation}\label{eq:Sup_Unitary2}
 U(s,g)=\prod_{t_j=1}^{N}e^{-idt\bigg[(1-t_j/T)H_{TT}+\frac{t_j}{T}H(g)\bigg]}   
\end{equation}
We verify that the unitary in Eq.~\eqref{eq:Sup_Unitary2} gives greater than $0.99999$ overlap with the exact GS of Eq.~\ref{eq:Sup_HYoshioka} for $L_y=12-14\ell_B$, where viscosity is quantized. 
The unitary can be implemented on quantum devices through Trotter-Suzuki and other techniques but will require early fault-tolerant machines. 
Hence, we leave the quantum implementation of this section for future work. 
We further verify that once GS of Eq.~\eqref{eq:Sup_HYoshioka} is obtained for $L_y>10\ell_B$ (for $g_{11}=g_{22}=1$), the state can be deformed adiabatically by applying the same unitary in Eq.~\eqref{eq:Sup_AdiabaticSE} for nondiagonal metric $g$ and repeating the same steps. 
Hence, both parts $\hat{U}_{1,\Delta_{12}}$ and $\hat{U}_{1+\Delta_{11},0}$ can be prepared for $L_y>10\ell_B$ and measured though the Hadamard test of the main text. 
Fault-tolerant implementation will not need modified Hadamard test.

\bibliography{references.bib}

\end{document}